# Integrated Maintenance: analysis and perspective of innovation in railway sector

Dr. Eng. Roberto Nappi
Chief Technology Officer
T&T s.r.l., Naples, Italy
roberto.nappi@ttsolutions.it

*Abstract*—The high costs for the management of the modern and complex industrial control systems make it necessary to enhance the current maintenance processes. Therefore, the need arises to clearly define the goals of the maintenance, in order to evolve and continuously enhance the management methods, to efficiently integrate the maintenance activities with the ones related to the production, the service provisioning, and the operation, and to use smart *computer-based* maintenance systems. This paper proposes a general maintenance approach and its specific application to the railway sector.

## INTRODUCTION

Many complex systems, in different engineering application fields (e.g. aerospace, aeronautic, naval, railway, etc.), work in specific environmental conditions for which it is required to be compliant with specific requirements of usability, reliability, safety, and maintainability. In particular, regarding these requirements, the target of the maintainability is to maximize the lifetime of the systems produced with the minimum global cost (*Life Cost Cycle*). Based on this consideration, the maintenance of a system becomes a strategic element for the economic competitiveness of the infrastructure operators . Indeed, in order to have a system of high complexity that properly operates, without interruptions, it is necessary to sustain its usage by a constant maintenance activity. The "traditional" approach to the maintenance is typically intended as repair. However, it is possible to evolve this approach through actions such as the prevention and the continuous improvement of the maintenance process focused on the system lifecycle (Fig. 1).

Fig. 1. Engineering Maintenance

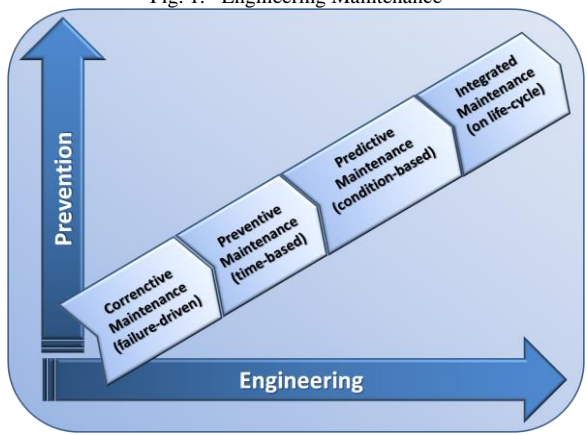

Through this new kind of management, the maintenance is applied more efficiently and is transformed in an innovative and competitive process.

The fundamental element for this evolution is the support given by the new technologies for the information transmission and management. The usage of these instruments allows to record and store the inspection and intervention data, thus enabling to process them with models and algorithms for the extraction of knowledge helpful to the continuous enhancement of maintenance plans and company's economic results. In this way, it will be possible to model the variables associated to the safe operation of the system and the ones related to the costs, in order to detect critical issues such as a combination of event's occurrences and their impact. Thus, using this "modern" approach it is possible to integrate technical and engineering aspects with economic and managerial aspects (Fig. 2).

Fig. 2. Integrated Maintenance (Technical and Engineering with Managerial and Economic aspects)

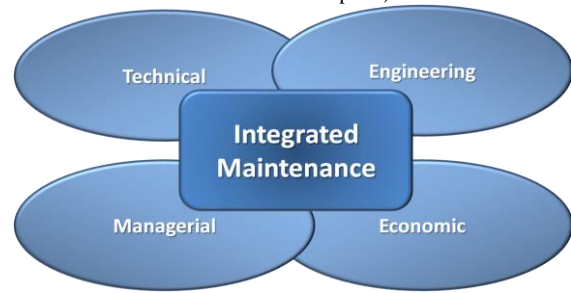

The high costs for the management of the modern and complex industrial control systems make it necessary to enhance the current maintenance processes. Therefore, the need arises to clearly define the goals of the maintenance, in order to evolve and continuously enhance the management methods, to efficiently integrate the maintenance activities with the ones related to the production, the service provisioning, and the operation, and to use smart *computer-based* maintenance systems to manage data from multiple sources. To deal with big data analysis, innovative algorithms and data mining tools are needed in order to extract information and discover knowledge from the continuous and increasing data volume. In most of data mining methods the data volume and variety directly impact the computational load. The centralized management of huge amount of data for complex maintenance activities implies to produce a system (i.e. a Critical Infrastructure), resources and processes, whose destruction or disruption, even partially, may, directly or indirectly, strongly to affect the normal, safe, secure and efficient functioning of a country (e.g., transportation, financial, power grid, etc.).

The paper is structured in the following sections. Section I presents maintenance approaches, while Section II provides an overview of maintenance standardization efforts. Section III illustrates perspectives of innovation in the management of maintenance in railway sector. Finally, Section IV gives some concluding remarks.

## I. MAINTENANCE APPROACHES

### A. Corrective Maintenance

The *failure-driven* approach is a reactive management approach, where the corrective maintenance is often dominated by unplanned events and it is performed only after the occurrence of failures or breakages of the system. Corrective Maintenance actions can recover the malfunctioning part of the system, repairing or replacing the failed component. If the system is not-critical and easily repairable, any potential unplanned crashes will cause a minimum impact related to the availability. In this way, the *failure-driven* maintenance can be a good maintenance approach. However, in case of purely random failures of the system, that could have a serious impact on the productivity, an urgent corrective maintenance action is required to avoid serious consequences produced by the fault. Therefore, the systematic use of an urgent corrective maintenance often is translated into unpredictable performance of the system, i.e. high *down-time* of the system, high costs for the recovering of system's functionalities, extended repair times, high penalties related to the system unavailability, an high level of stocks for the replacement parts.

### B. Preventive Maintenance

The *time-based* maintenance is also known as periodic preventive maintenance. In order to slow down the process of deterioration that leads to a failure, a primary preventive maintenance is performed periodically inspecting and controlling the system through scheduled regular activities. The *time-based* maintenance assumes that the estimated malfunctioning of the system, i.e. mean time between two functional failures (*Mean Time Between Failure* – MTBF), is statistically or experimentally known for system and device degradation during their normal use. The *time-based* maintenance involves also scheduled shutdown of the system for revisions or predetermined repair activities on the system still operating. This approach allows to prevent functional failures thanks to the replacing of critical components at regular intervals, just shorter than their estimated life cycle. The system revision and the replacement of critical components at determined intervals represent methodologies widely adopted in the maintenance of many modern systems. Although the *time-based* maintenance can reduce the failure probability of a system or the frequency of unplanned emergency repairs, it cannot delete the occurrences of random failures. Some *time-based* maintenance practices may be obsolete and unable to cope with the current operational requirements of modern automated systems. The maintenance decisions are made by experienced planners, according to the: recommendations of the manufacturer of the system, failures history, malfunction data, operational experience, assessment performed by maintenance staff and technicians. Vice versa, under circumstances of uncertainty resulting from random failures, it is difficult to properly plan maintenance activities in advance and, consequently, the maintenance staff is forced to operate in emergency conditions.

### C. Predictive Maintenance

Another approach is represented by the *condition-based* maintenance as a method to reduce the uncertainty of maintenance activities. These activities will be performed according to the needs indicated by the results of system status monitoring (*condition-monitoring*). The predictive *condition-based* maintenance uses, therefore, the results of *condition-monitoring* and, according to these, plans the maintenance actions. The goal of *condition-monitoring* is to delete the failures and extend the preventive maintenance intervals. The *condition-based* maintenance assumes that the existence of indicative prognostic parameters can be identified and used to quantify potential system failures before they occur. The prognostic parameters provide an indication of potential problems and new issues that may cause the deviation of the system from its acceptable level of functioning. The *condition-based* fault diagnosis is triggered by the detection of an evaluated condition of the system, such as the deviation from the expected level, recognizes and analyzes symptomatic information, identifies the causes of the malfunction, obtains the development trend of the fault and predicts the remaining useful life of the system (*Remaining Useful Life* – RUL). In order to obtain a fully automated system for *condition-monitoring*, new analysis techniques need to be used, such as Artificial Intelligence, able to handle large amounts of data, Neural Networks, Motivation Case-Based and Fuzzy Logic. Equipped with this such predictive skill, the diagnostic system becomes more reliable. Once a component has been identified as the cause of a new failure, the function devoted to prediction of the development trend of the fault can be activated to compute the remaining life time. Furthermore a scheduled action of corrective maintenance can be appropriately performed before a possible irreversible damage of the system occurs. A maintenance in advance can be performed in order to avoid an excessive supply of replacement parts. Therefore, the implementation of an automated condition-monitoring process provides a better and timely determination of the maintenance interventions, which will result in a decrease of the life cost of the system, thanks to an increment of its availability and to a reduction of operations and maintenance costs.

### D. Integrated Maintenance

For the most modern and complex industrial systems, the attention of the manager of maintenance shall be focused on the following three aspects:

- how to re-plan and preschedule the maintenance of sophisticated systems operating in complex environmental conditions;

- how to reduce the high costs of stocks of the replacement parts;

- how to avoid risks of catastrophic failures and eliminate the forced and unplanned interruption in system availability.

A Decision Support System (*Decision Support System – DSS*) is a *computer-based* processing system that contains a specific knowledge of domain and analytic decisional models to assist the decision-maker through the presentation of information and interpretation of possible operational alternatives. The decision support system is aimed to improve the decision-making process of maintainer, providing an easy definition and identification of the problem, an appropriate management of information and statistical tools, with a proper application of the knowledge. It is designed to enable a quick and easy generation of alternatives and to increase awareness of weaknesses in the decision-making process. It can help the decision maker to adopt more efficient and effective strategies in complex situations. Therefore, the *condition-based* fault diagnosis and the prediction of the deterioration of the system are essential into maintenance management approaches. In order to provide a complete tool for decision support in maintenance management, it is required to proceed towards the integration of *condition-monitoring*, intelligent diagnosis of the *condition-based* faults and prediction of deterioration trend of the system. All this encourages the development of a new technology sector capable of offering an intelligent system as support tool to predictive decisions, that includes the prediction of deterioration trend of the system for a *condition-based* maintenance. Therefore, an *Intelligent Predictive Decision Support System* (IPDSS), for a *condition-based* maintenance, integrates the following concepts:

- monitoring of system condition;
- intelligent faults diagnosis, based on system condition;
- prediction of deterioration trend of the system.

Through the integration of these three elements, the quality of maintenance will dramatically improve (*Intelligent System* = System able to autonomously make decisions). Fig. 3 shows the operations of a IPDSS system that provides maintenance alerts depending on the state of the system.

Fig. 3. Management of maintenance activities through an Intelligent Predictive Decision Support System (RUL=Remaining Useful Life)

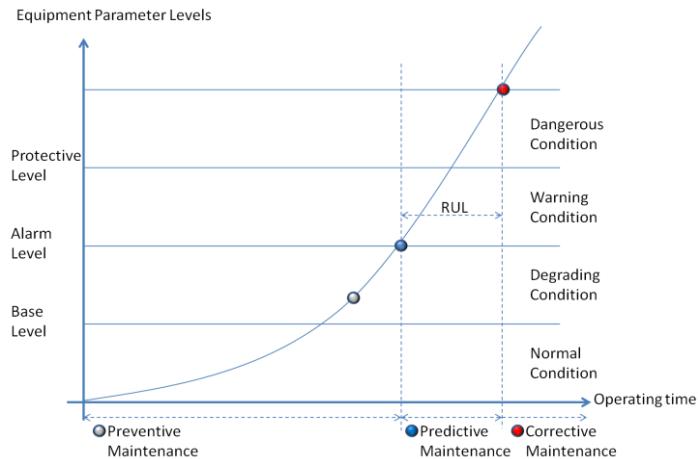

When the system operates in "normal condition", only a generic preventive maintenance activity is required. Instead, when the monitored parameters of the system reach the "base level", the system goes into "degraded condition" and this indicates that failures may be experienced. In this case, the analysis of the development trend of faults should be conducted in order to indicate the areas of the system affected by potential problems due to new faults or areas where it is probable that faults occur. When a "degraded condition" occurs, no special maintenance action is required; instead, more *condition-monitoring* actions should be performed in order to prevent emergency interventions. When the monitored parameters of the system exceed the "alarm level", then appropriate indications alarm will be activated in order to alert the operators and the maintenance staff. The *case-based* diagnostic system of IPDSS will be subsequently activated to search for a similar situation within the historic maintenance cases. If a system failure already archived occur, then the *case-based* diagnostic system will elevate the maintenance notice reporting the problem-cause-solution according to the historical of the maintenance cases. Vice versa, when the detected condition is not present into the *well-known* maintenance cases, the IPDSS will activate the *rule-based* or *model-based* system for failure diagnosis. The *rule-based* system can detect and identify new faults in accordance with the rules representing the relations between every possible failure and the current condition of the monitored system. The *model-based* diagnostic system enhances the diagnosis capability based on the structure and the properties of the system, which includes logical and mathematical methods. Once a component has been diagnosed as the cause of a future malfunction, the function of prediction of system deterioration can be activated to evaluate the remaining life time of the faulty system. The result of the prediction can later be used to organize the corrective maintenance action, planned before the system failure as well as to avoid possible irreversible damages. When the parameters of the monitored system reach the "predicted level", the system will produce a sudden or almost complete stop as result of a system failure. Under this danger condition, an emergency corrective maintenance action will be started immediately.

## II. MAINTENANCE STANDARDIZATION

In the industrial field high costs related to maintenance activities are yearly expected planned. The high costs are related to the loss of productivity due to a poor availability of assets regularly maintained. Using a *condition-based* maintenance, the maintenance intervals will be determined on dynamic conditions and no activities will be performed when not needed. The current progress in sensor technologies make it possible to monitor the most crucial components of a system. In addition, different techniques of artificial intelligence give the possibility to analyze the measured data. With the knowledge of the process of the monitored asset, the analysis will show if the maintenance actions originally planned need to be improved or not. The *condition-based* approach has been successfully used in different industrial sectors and, when correctly applied, it allows to save up to 20% in stocks of replacement parts, to reduce losses of production and quality. Standardization efforts in the area of integrated system maintenance goes along three lines:

- a standard for an easy transduction interface for sensors and actuators (IEEE 1451);

- a proposal of standardization for the architecture of condition-monitoring systems (OSA-CBM);
- a proposal of standardization for the communication between different condition-monitoring systems (MIMOSA).

Through these standards, potentially jointly applicable, the interchangeability of hardware and software components can be achieved, giving more technological options to users, more rapid technological development, price reduction and an easier update of the system components. When the developers of maintenance techniques based on *condition-monitoring* apply standardized integration processes, it is easier to drive the development of algorithms and new methods for predicting the remaining useful lifetime of the system components. In order to lead the development process towards a more general maintenance system *condition-based* is important to proceed with the use of standards, contributing also to the progress of existing proposals.

*A. IEEE 1451*

Due to the problems encountered by users during the activities of products integration (transducer, sensors and actuators) of different vendors and their network connection, it is necessary to adopt a standard for the hardware and software interconnection level, in order to obtain the interoperability in the exchange and in the use of information. Over time, the development industries and the ones that use distributed measurement and control systems have moved away from proprietary standards and are oriented towards systems with "open" approaches, standard *de facto*. In any case, trying to develop a standard interface for intelligent sensors, the *National Institute of Standards and Technology* (NIST), in cooperation with the *Institute of Electrical and Electronics Engineers* (IEEE), has started to work on this objective since the mid of 90's. This purpose is subsequently became the standard IEEE 1451, which aims to achieve common interfaces in order to connect transducers towards systems based on microprocessors and towards tools and field networks, avoiding that the operation related to a network node (insertion/deletion) can influence the behavior of the other nodes.

*B. OSA-CBM*

OSA-CBM is the acronym of *Open System Architecture for Condition Based Maintenance* and is a proposal for a standard *de-facto* non-proprietary. The mission of OSA-CBM organization states that the standard proposal should cover the entire range of functions of a *condition-based* maintenance system, both for hardware and software components. The proposed standard divides the *condition-based* maintenance system in seven different levels, all interconnected (Figure 4). <u>Level 1 Sensor Module</u>. It provides sensors that return digitalized results or transducers that return data. Signal module could be built following the standard IEEE 1451. <u>Level 2: Signal Processing</u>. The module receives signals and data from the sensor module or other modules of signal processing. The output of signal processing module includes sensor-data digitally filtered, frequency spectrum, signals of virtual sensor and other features related to the condition-based maintenance.

The signal processing module may consist of an AI-ESTATE (*Artificial Intelligence and Expert System Tie to Automatic Test Equipment*), as reported in IEEE 1232 standard. <u>Level 3 Condition Monitor</u>. The condition-monitor level receives data from sensor modules, signal processing modules and other *condition-monitor* modules. The main goal of this level is to compare data with their expected values. For example, regarding the vibrations, ISO 13373 may be used as reference. The condition-monitor level shall be also able to generate alerts based on operational limits previously set. This latter can be a very useful function during development of rapid failures. <u>Level 4: Health Assessment</u>. The module devoted to the assessment of the "status of health" receives data from different *condition-monitor* modules or other modules of assessment of the system conditions. The main goal of the condition assessment module is to determine if the condition of the monitored component/subsystem/system is degraded. The evaluation module shall be able to generate diagnostic recordings and propose failure estimation. The diagnosis shall be based on trends of the health status history, on operating status, workload and maintenance history. <u>Level 5: Prognostics</u>. The prognostic module shall be able to take into account data from all the previous levels. The main goal of the prognostic module is to compute the future health status of an asset, taking into account its future profile of usage. The module will report the future health status at a specified time or, alternatively, the remaining useful lifetime. <u>Level 6: Decision Support</u>. The decision support module receives data from the module of health status evaluation and the prognostic module. Its main goal is to generate the recommended actions and the alternatives ones. Actions may be of maintenance type but also related to how to run an asset until the current mission is completed without the occurrence of breakage. <u>Level 7: Presentation</u>. The presentation module must show the data coming from all the previous modules. The most important levels of which present the data are those related to Health Assessment, Prognostic and Decision Support, as well as the alarms generated by the condition-monitor modules. The presentation module can also have the opportunity to look further downwards and can be inserted also into a machine-interface.

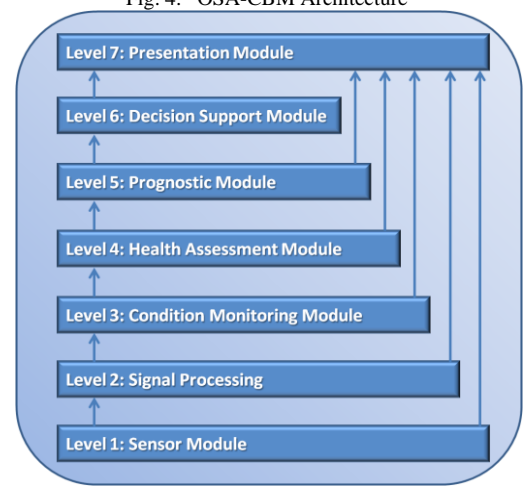

Fig. 4. OSA-CBM Architecture

*C. MIMOSA*

The *Machinery Information Management Open System Alliance* (MIMOSA) was founded in 1994. Its purpose and goal is to develop open conventions for the exchange of information between the computer systems of maintenance and installation. The development of MIMOSA CRIS (*Common Relational Information Schema*) has been published and covers data and information that will be managed within a *condition-monitoring* maintenance system.

### III. Perspectives of innovation in the management of maintenance in railway sector

*A. Analysis*

The railways have always been one of the major development factor for the modern industrial society, having a considerable impact on their processes of economic growth. They represent an efficient for the transportation of small and large quantities of passengers and freights, over short and long distances. The need to have adequate transportation, therefore, characterizes the modern societies contributing to the improvement of life quality. For the most globalized industrial societies, the comparison between different systems of railway transport appears natural and is addressed to assess overall efficiency in the management of services provided. Therefore, the overall improvement of a railway transport system represents the new technological challenge for this industrial sector and the maintenance strategies represent a fundamental aspect of this challenge. After several comparison made between the main railways operators analyzed (railway infrastructure managers, trains fleets managers), some differences were found both in the management of operating costs than in the policies of management of maintenance activities. These two elements, if compared, show that the operators that adopt maintenance policies based on system status (*condition-based maintenance*) support overall management costs lower than the operators who, instead, adopt only maintenance policies based on the operating time of the components (*time-based maintenance*).

In the railway sector, the maintenance intervals are typically determined "statistically", because they are based on operating time and/or distance traveled, or on the amount of actions performed by the system. These intervals are based on previous experiences or on the specification made by railways operators and they are chosen, mainly, on the basis of the average life of the components involved. Nevertheless, it is possible to have components that deteriorate more rapidly than expected, for example as a result of the influence of specific external factors. Therefore, the traditional maintenance method can be enhanced when the degraded variations of system status, due to wear of exercise, can be appropriately monitored. An evolution of maintenance management, then based on the current status of the system and on a "dynamic" identification of maintenance intervals, would lead to detect unexpected deteriorations of the components, thus increasing the availability, the reliability and the safety of the railway, and contributing to economic optimization of their use.

Next step in the evolution of maintenance in railway sector leads, therefore, toward a more preventive approach, based on system condition and on the prediction of the evolution of its state. Follow the status of the system makes it possible, therefore, its most efficient use, because the maintenance activities can be planned in advance and with greater accuracy, advantaging the availability of the system, which can be managed accordingly. In addition, the *condition-based* maintenance approach contributes to the reduction of operative costs and to reduction of the level of operative support, during the useful life of the system. In order to implement this maintenance policy is necessary, therefore, to have different type of monitoring the state of the system (*condition-monitoring*) and be able to predict the remaining lifetime of its components, analyzing and producing forecast trend starting from the measures performed. It becomes, therefore, essential to have a technological infrastructure based on the collecting, organizing and managing of data.

*B. State of the art*

Generally, the current tools for observing the elements of a railway system work separately, not having an integrated vision and a joint use of the acquired data. Starting from the analysis made, it is possible to classify these tools as follow:

- <u>tools for the monitoring of the railway infrastructure installed on the infrastructure</u>, able to observe its status. Some of these tools are able, starting from the monitoring of railway infrastructure status, to trace back the deterioration phenomena of rolling stock. Indeed, the *condition-monitoring* applied to railway infrastructure is a powerful tool used both to control its status and the status of the vehicles using, as point of view, the railway infrastructure itself. Monitoring the condition of the infrastructure can help also to prevent problems in opposite direction, i.e. as the infrastructure condition influences the status of the rolling stocks. Indeed, a failure on the infrastructure may leads to a failure on the rolling stock which may have, in its turn, impact on the infrastructure after the traveling of the train along the line;

- <u>tools for the monitoring of the railway infrastructure installed on rolling stocks</u>, i.e. installed on dedicated diagnostics trains or service train, correlating the measurements made with the information of spatial localization. In this way, it is possible to obtain an overview of the infrastructure condition, controlling the status in order to evaluate possible degradations, identifying new faults and railways line areas interested by problems. In addiction to wheel/rail interface, for the electrified railways the interaction between the contact line and the pantograph is a monitored interface. A failure on the interface wheel/rail can leads to a derailment, causing victims, economic and material damages; at the same time, a failure on the interface between contact line and pantograph, as well as representing a risk for potential physical damages, can leads to availability problems of long duration in the management of railway traffic and it can be source of significant costs. On the market, there are different monitoring systems able to evaluate the effects of pantograph degradation on the infrastructure and, vice versa, evaluate the effects of contact line degradation on the pantograph;

- tools for the monitoring of the rolling stocks status installed on rolling stocks, able to detect the degradation of different mechanics and electrics onboard parameters, with relative predictive evaluations of the status of the monitored components.

The availability of all these tools and the technology progress in the monitoring of system status, show that the *condition-based* maintenance strategy, managed in a manner that integrates all the point of view, will be able to become a powerful instrument to reduce operating and maintenance costs. The information obtained by different measurements, actually disaggregated and heterogeneous, will be able to be integrated and elaborated homogeneously by predictive models and innovative expert systems, in order to support the management decision. These tools will be further developed and refined over time, thanks to a growing increase of the investments in the railway industry.

One of the challenges in the implementation of *condition-monitoring* is to find the measurements technologies suited to monitoring the status of the specific system, able to provide valid and reliable measures. Physical sensors are adopted to monitor the operating environment conditions: monitoring some parameters is necessary to guarantee safe execution of critical processes, avoid hazardous controls and prevent possible critical failures or damages to the observed infrastructure. To monitor environmental parameters, several instruments and sensor devices exist. With regards to the most advanced technologies, the smart sensors can be mentioned. These devices have increased the capabilities of the metering process in several aspects. Indeed, smart sensors have introduced the possibility to process the measures on the sensor boards, sending alarm messages in case of suspect environmental conditions on event detection. Other kinds of smart devices, such as the Sensor Networks, have provided capabilities in terms of protection mechanisms able to isolate faulted and misbehaving nodes. It is also necessary to identify the parameters to be measured able to provide data essential for the production of relevant and acceptable information, to be used as support to decision in the process of maintenance management. The technical issues related to the data booming have been designated as the big data challenges and have been identified as highly strategic by major research agencies. Most definitions of big data refer on the so-called three "V"s: volume, variety and velocity, referring respectively to the size of data storage, to the variety of source and to the frequency of the data generation and delivery. To deal with big data analysis, innovative approaches for data mining and processing are required in order to enable process optimization and enhance decision making tasks. To achieve this, an increment on computational power is needed and dedicated hardware can be adopted. There are two main classes of solutions: 1) using general purpose CPUs as multi-core processors and/or computer clusters to run the data mining software; 2) using dedicated hardware (special purpose) to compute specific parts of an algorithm, reducing the computational effort. Indeed special purpose machines may not be suitable as they are not programmable and many classification systems need tuning and reprogramming to achieve high accuracy. Nevertheless Field Programmable Gate Array (FPGA) can be adopted as Decision Support System for the low costs, the easy re-configuration, the reliability and the safety related properties useful in the real-time embedded systems used in the railway sector.

IV. CONCLUSIONS

The *condition-based* maintenance strategy, if properly used, represent an element of absolute efficiency improvement compared to the use of the only *time-based* and *failure-driven* strategies. It is important to note the importance of integration between different tools of *condition-monitoring*, which complement each other in the measurement of different parameters, and the importance of integrate these tools also considering pre-existing maintenance processes and tools, with the aim of exploiting all the benefits arising from the application of new technologies in predictive optic, in order to support the most appropriate maintenance decisions to be undertaken. It is fundamental the opportunity to measure the status in a continuous way, correct and accepted by all stakeholders. In fact, it is important the sharing of this approach between the manager of railway infrastructure, the manager of rolling stocks and the maintenance companies, obtaining in this way a significant impact on the overall efficiency of railway exercise. In order to achieve this goal, it will be necessary to have tools (technics and economics) that will support:

- the overall monitoring of the different technological elements of a railway system;
- the predictive assessments on the elements status of the railway system;
- predictive models of costs;
- management of the choices between the different stakeholders.

In this context, the use of standardized technologies, or the participation in standardization processes, helps the management of maintenance according to an integration strategy. In fact, the integrated maintenance approach, addressing technological and engineering aspects, as well as managerial and financial aspects, allows to include the maintenance at all stages of the life cycle of the designed system. Moreover, the current standardization processes can also include aspects related to Critical Infrastructure Protection to prevent misuses and malicious activities causing catastrophic consequences, such as danger for life and financial losses, when there is need for a complex and centralized big data management during the integrated maintenance activities.

The simultaneous application of the same criteria to the maintenance management, both for aspects related to vehicles and for those related to railway infrastructure, would allow to obtain significant improvements, thus implementing a complete process of collection and centralized management of information and maintenance actions. Therefore, partnerships between all actors involved in this joint process represent a feasible perspective, considering the heterogeneity characterizing the tasks of the manager of railway infrastructure, the manager of rolling stocks and the manager of the maintenance.